\documentclass[sigconf]{acmart}
\usepackage{booktabs}
\usepackage{siunitx}
\usepackage{caption}
\usepackage{subfigure}
\usepackage{amssymb}
\usepackage{float}

\AtBeginDocument{%
  \providecommand\BibTeX{{%
    \normalfont B\kern-0.5em{\scshape i\kern-0.25em b}\kern-0.8em\TeX}}}

\setcopyright{acmcopyright}
\copyrightyear{2022}
\acmYear{2022}

\acmConference[ORSUM @ ACM RecSys]{ORSUM@ACM Recsys 2022}{September 23, 2022}{Seattle, WA}

\begin{document}

\title{On the Factory Floor: ML Engineering for \\ Industrial-Scale Ads Recommendation Models}

\author{Rohan Anil, Sandra Gadanho, Da Huang, Nijith Jacob, Zhuoshu Li, Dong Lin,\\ Todd Phillips, Cristina Pop, Kevin Regan, Gil I. Shamir, Rakesh Shivanna, Qiqi Yan}
\affiliation{%
  \institution{Google Inc.}
  \country{}
}

\settopmatter{printacmref=false}

\renewcommand{\shortauthors}{Anil, et al.}

\begin{abstract}
For industrial-scale advertising systems, prediction of ad click-through rate (CTR) is a central problem. Ad clicks constitute a significant class of user engagements and are often used as the primary signal for the usefulness of ads to users. Additionally, in cost-per-click advertising systems where advertisers are charged per click, click rate expectations feed directly into value estimation. Accordingly, CTR model development is a significant investment for most Internet advertising companies. Engineering for such problems requires many machine learning (ML) techniques suited to online learning that go well beyond traditional accuracy improvements, especially concerning efficiency, reproducibility, calibration, credit attribution. We present a case study of practical techniques deployed in Google's search ads CTR model. This paper provides an industry case study highlighting important areas of current ML research and  illustrating how impactful new ML methods are evaluated and made useful in a large-scale industrial setting.
\end{abstract}

\maketitle

\section{Introduction}
\label{sec:intro}

Ad click-through rate (CTR) prediction is a key component of online advertising systems that has a direct impact on revenue, and continues to be an area of active research 
\cite{mcmahan2013ad,xinran2014FB,zhou2018deep,ling2017model}.
This paper presents a detailed case study to give the reader a "tour of the factory floor" of a production CTR prediction system, describing challenges specific to this category of large industrial ML systems and highlighting techniques that have proven to work well in practice.

The production CTR prediction model consists of billions of weights, trains on more than one hundred billion examples, and is required to perform inference at well over one hundred thousand requests per second. The techniques described here balance accuracy improvements with training and serving costs, without adding undue complexity: the model is the target of sustained and substantial R\&D and must allow for effectively building on top of what came before.

\subsection{CTR for Search Ads Recommendations}

The recommender problem surfaces a result or set of results from a given corpus, for a given initial context. The initial context may be a user demographic, previously-viewed video, search query, or other. Search advertising specifically looks at matching a \textbf{query} $q$ with an \textbf{ad} $a$. CTR models for recommendation specifically aim to predict the probability $P(click|x)$, where the input $x$ is an ad-query pair $\langle a, q \rangle$, potentially adorned with additional factors affecting CTR, especially related to \textbf{user interface}: how ads will be positioned and rendered on a results page (Section \ref{sec:uinorms}). 

Beyond surfacing maximally useful results, recommender systems for ads have important additional \textbf{calibration} requirements. 
Actual click labels are stochastic, reflecting noisy responses from users. For any given ad-query $x_i$ and binary label $y_i$, we typically hope to achieve precisely $P(click|x_i) := \mathbb{E}_{\langle x_i, y_i \rangle \sim D}[y_i=click| x_i]$ over some sample of examples $D$ (in test or training). While a typical log-likelihood objective in supervised training will result in zero aggregate calibration bias across a validation set, per-example bias is often non-zero.

Ads pricing and allocation problems create the per-example calibration requirement. Typically, predictions will flow through to an auction mechanism that incorporates bids to determine advertiser pricing. Auction pricing schemes (e.g, VCG \cite{varian2014vcg}) rely on the relative value of various potential outcomes. This requires that predictions for all potential choices of $x$ be well calibrated with respect to each other. Additionally, unlike simple recommenders, ads systems frequently opt to show no ads. This requires estimating the value of individual ads relative to this "null-set" of no ads, rather than simply maximizing for ad relevance.

Consider a query like "yarn for sale"; estimated CTR for an ad from "yarn-site-1.com" might be 15.3\%. Estimated CTR for an ad from "yarn-site-2.com" might be 10.4\%. 
Though such estimates can be informed by the semantic relevance of the websites, the requirements for precision are more than what one should expect from general models of language. Additionally, click-through data is highly non-stationary: click prediction is fundamentally an online recommendation problem. An expectation of 15.3\% is not static ground truth in the same sense as, for example, translation or image recommendation; it is definitively more subject to evolution over time.

\subsection{Outline}

For ads CTR predictors, minor improvements to model quality will often translate into improved user experience and overall ads system gains. This motivates continuous investments in model research and development. Theoretical and benchmark improvements from ML literature rarely transfer directly to problem-dependent settings of real-world applications. As such, model research must be primarily empirical and experimental. Consequently, a great deal of attention must be paid to the machine costs of model training experiments while evaluating new techniques. In Section~\ref{sec:systems} we first give a general overview of the model and training setup; Section~\ref{sec:efficiency} then
discusses \textbf{efficiency} concerns and details several successfully deployed techniques. In Section~\ref{sec:accuracy}, we survey applications of modern ML techniques targeted at improving measures of {\bf accuracy} and geared explicitly toward very-large-scale models. Section~\ref{sec:summary-eff-acc} summarizes empirical results roughly characterizing the relative impact of these techniques. 

Deep neural networks (DNNs) provide substantial improvements over previous methods in many applications, including large-scale industry settings. However, non-convex optimization reveals (and exacerbates) a critical problem of prediction: \textbf{irreproducibility} \citep{ dusenberry20,shamir20a,shamir20,shamir2020smooth,snapp2021synthesizing, damour20}.
Training the same model twice (identical architecture, hyper-parameters, training data) may lead to metrics of the second model being very different from the first.
We distinguish between {\bf model irreproducibility}, strictly related to predictions on fixed data, and {\bf system irreproducibility}, where a deployed irreproducible model affects important system metrics. Section~\ref{sec:reproducibility} characterizes the problem and describes improvements to model irreproducibility. 

An effective click prediction model must be able to {\bf generalize across different UI treatments}, including: where an ad is shown on the page and any changes to the formatting of the ad (e.g., bolding specific text or adding an image). Section~\ref{sec:uinorms} describes a specific model factorization that improves UI generalization performance. Finally, Section~\ref{sec:copt} details a general-purpose technique for adding {\bf bias constraints} to the model that has been applied to both improve generalization and system irreproducibility. 

This paper makes the following contributions:
1) we discuss practical ML considerations from many perspectives including accuracy, efficiency and reproducibility,
2) we detail the real-world application of techniques that have improved efficiency and accuracy, in some cases describing adaptations specific to online learning, and 
3) we describe how models can better generalize across UI treatments through model factorization and bias constraints.

\section{Model and Training Overview} \label{sec:systems}

A major design choice is how to represent an ad-query pair $x$. The semantic information in the language of the query and the ad headlines is the most critical component. Usage of attention layers on top of raw text tokens may generate the most useful language embeddings in current literature \citep{vaswani2017attention}, but we find better accuracy and efficiency trade-offs by combining variations of fully-connected DNNs with simple feature generation such as bi-grams and n-grams on sub-word units. The short nature of user queries and ad headlines is a contributing factor. Data is highly sparse for these features, with typically only a tiny fraction of non-zero feature values per example.

All features are treated as categorical and mapped to sparse embedding tables. Given an input $x$, we concatenate the embedding values for all features to form a vector $e$, the \textbf{embedding input layer} of our DNN. $E$ denotes a minibatch of embedding values $e$ across several examples. 

Next, we formally describe a simplified version of the model's fully-connected neural network architecture. Later sections will introduce variations to this architecture that improve accuracy, efficiency, or reproducibility. We feed $E$ into a fully-connected hidden layer $H_1\!=\!\sigma(E W_1)$ that performs a linear transformation of $E$ using weights $W_1$ followed by non-linear activation $\sigma$. Hidden layers $H_i\!=\!\sigma(H_{i-1} W_i)$ are stacked, with the output of the $k$th layer feeding into an output layer $\hat{y}\!=\! \text{sigmoid}(H_k W_{k+1})$ that generates the model's prediction corresponding to a click estimate $\hat{y}$. Model weights are optimized following $\min_{W} \sum_{i} \mathcal{L}(y_i, \hat{y}_i)$. We found ReLUs to be a good choice for the activation function; Section~\ref{sec:reproducibility} describes improvements using \emph{smoothed} activation functions. 
The model is trained through supervised learning with the logistic loss of the observed click label $y$ with respect to $\hat{y}$. Sections~\ref{sec:accuracy} and \ref{sec:copt} describe additional losses that have improved our model. Training uses synchronous minibatch SGD on Tensor Processing Units (TPUs) \citep{jouppi2020domain}: at each training step $t$, compute gradients $G_t$ of the loss on a batch of examples (ranging up to millions of examples), and weights are optimized with an adaptive optimizer. We find that AdaGrad \cite{mcmahan2010adaptive,duchi2011adaptive} works well for optimizing both embedding weights and dense network weights. Moreover, In Section~\ref{sec:secondorder} discusses accuracy improvements from deploying a \textbf{second-order optimizer}: \mbox{Distributed Shampoo} \cite{anil2020shampoo} for training dense network weights, which to our knowledge, is the first known large-scale deployment in a production scale neural network training system.

\subsection{Online Optimization}
\label{sec:online}

Given the non-stationarity of data in ads optimization, we find that online learning methods perform best in practice \citep{mcmahan2013ad}. 
Models train using a single sequential pass over logged examples in chronological order. 
Each model continues to process new query-ad examples as data arrives \cite{swaminathan15batch}. For evaluation, we use models' predictions on each example from before the example is trained on (i.e., \textbf{progressive validation}) \citep{blum99}. 
This setup has a number of practical advantages. Since all metrics are computed before an example is trained on, we have an immediate measure of generalization that reflects our deployment setup. Because we do not need to maintain a holdout validation set, we can effectively use all data for training, leading to higher confidence measurements. This setup allows the entire learning platform to be implemented as a single-pass streaming algorithm, facilitating the use of large datasets.

\section{ML Efficiency} \label{sec:efficiency}

Our CTR prediction system provides predictions for all ads shown to users, scoring a large set of eligible ads for billions of queries per day and requiring support for inference at rates above 100,000 QPS. Any increase in compute used for inference directly translates into substantial additional deployment costs. Latency of inference is also critical for real-time CTR prediction and related auctions. As we evaluate improvements to our model, we carefully weigh any accuracy improvements against increases in inference cost. 

Model training costs are likewise important to consider. For continuous research with a fixed computational budget, the most important axes for measuring costs are bandwidth (number of models that can be trained concurrently), latency (end-to-end evaluation time for a new model), and throughput (models that can be trained per unit time).

Where inference and training costs may differ, several ML techniques are available to make trade-offs. Distillation is particularly useful for controlling inference costs or amortizing training costs (see Section~\ref{sec:distillation}). Techniques related to adaptive network growth \cite{chen2015net2net} can control training costs relative to a larger final model (with larger inference cost).

Efficient management of computational resources for ML training is implemented via maximizing model throughput, subject to constraints on minimum bandwidth and maximum training latency. We find that required bandwidth is most frequently governed by the number of researchers addressing a fixed task. For an impactful ads model, this may represent many dozens of engineers attempting incremental progress on a single modelling task. Allowable training latency is a function of researcher preference, varying from hours to weeks in practice. Varying parallelism (i.e., number of accelerator chips) in training controls development latency. As in many systems, lowered latency often comes at the expense of throughput. For example, using twice the number of chips speeds up training, but most often does so sub-linearly (training is less than twice as fast) because of parallelization overhead.

For any given ML advancement, immediate gains must be weighed against the long-term cost to future R\&D. For instance, naively scaling up the size of a large DNN might provide immediate accuracy but add prohibitive cost to future training (Table~\ref{tab:accuracy-summary} includes a comparison of techniques and includes one such naive scaling baseline). 

\begin{figure*}
\centering
\begin{minipage}{.34\textwidth}
  \centering
  \includegraphics[width=1.0\textwidth]{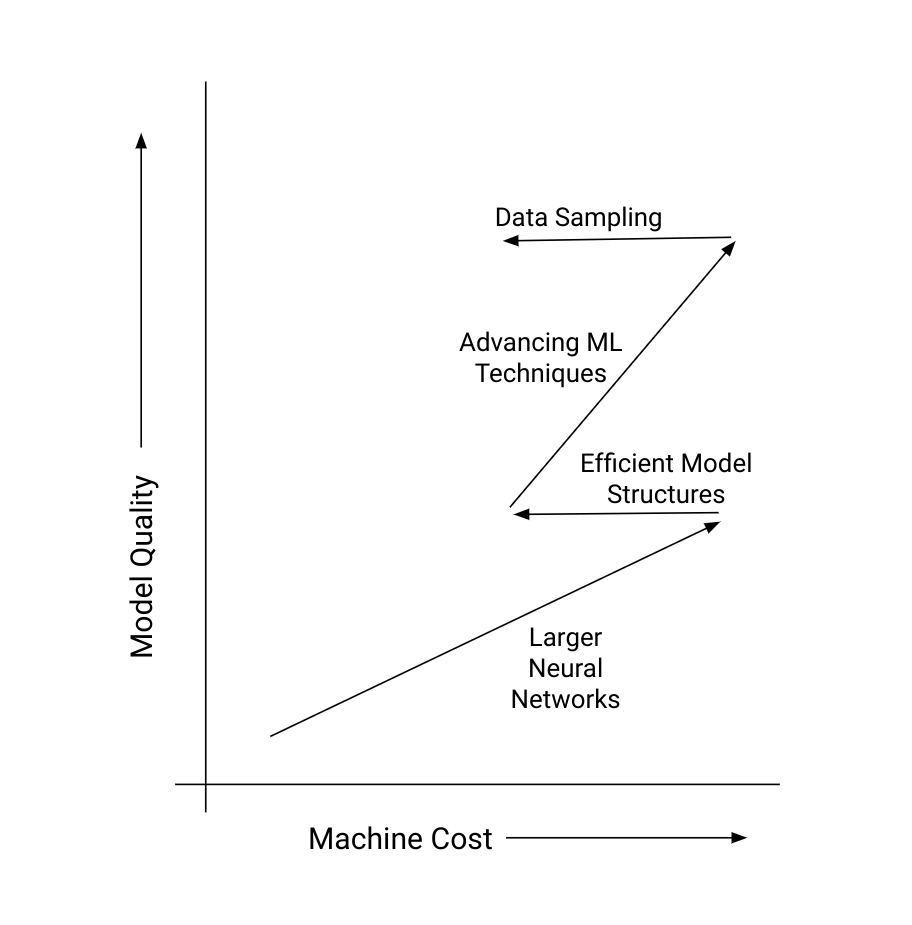}
  \captionof{figure}{\small "Switch-backs" of incremental costly quality-improving techniques and efficiency methods. \emph{(Illustration not to any scale.)}}
  \label{fig:cost-benefit}
\end{minipage}%
\begin{minipage}{.66\textwidth}
  \centering
  \includegraphics[width=0.8\textwidth]{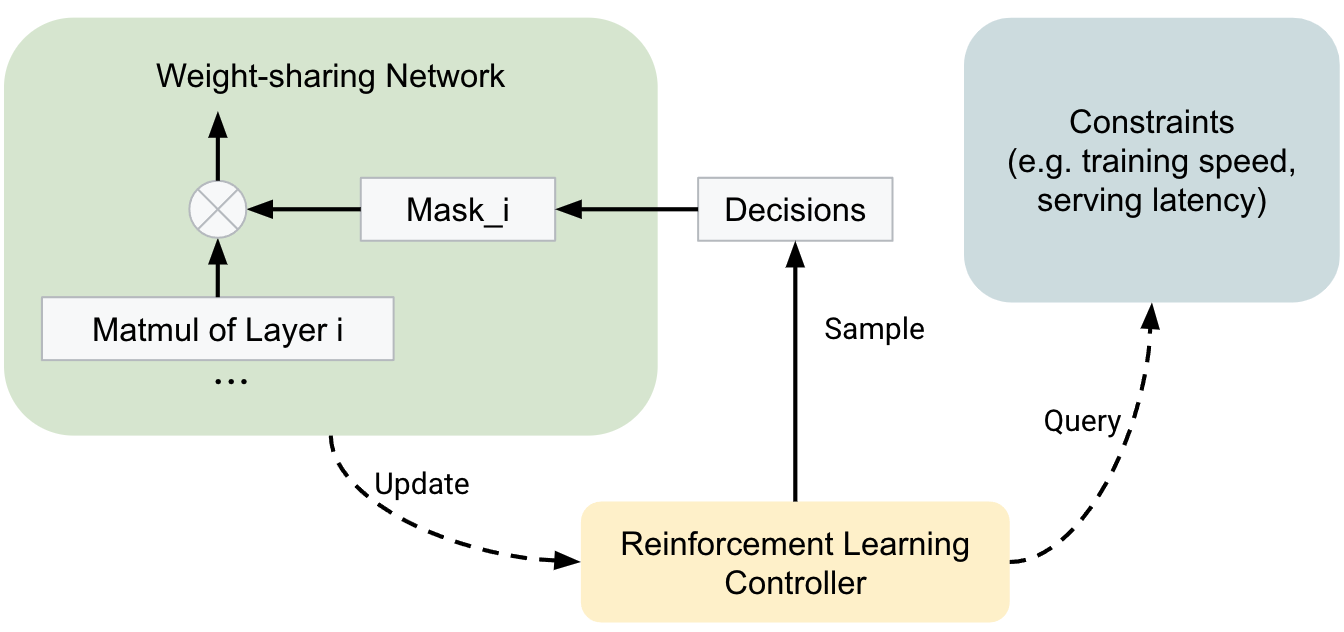}
  \captionof{figure}{\small Weight-sharing based NAS with cost constraints.}
  \label{fig:automl}
\end{minipage}
\end{figure*}

We have found that there are many techniques and model architectures from literature that offer significant improvements in model accuracy, but fail the test of whether these improvements are worth the trade-offs (e.g., ensembling many models, or full stochastic variational Bayesian inference \cite{blundell2015weight}). We have also found that many accuracy-improving ML techniques can be recast as efficiency-improving via adjusting model parameters (especially total number of weights) in order to lower training costs. Thus, when we evaluate a technique, we are often interested in two tuning points: 1) what is the improvement in accuracy when training cost is neutral and 2) what is the training cost improvement if model capacity is lowered until accuracy is neutral. In our setting, some techniques are much better at improving training costs (e.g., distillation in Section~\ref{sec:distillation}) while others are better at improving accuracy. Figure~\ref{fig:cost-benefit} illustrates these two tuning axes.

We survey some successfully deployed efficiency techniques in the remainder of this section. Section~\ref{sec:bottlenecks} details the use of matrix factorization bottlenecks to approximate large matrix multiplication with reduced cost. Section~\ref{sec:automl} describes AutoML, an efficient RL-based architecture search that is used to identify model configurations that balance cost and accuracy. Section~\ref{sec:sampling} discusses a set of effective sampling strategies to reduce data used for training without hurting accuracy.

\subsection{Bottlenecks} \label{sec:bottlenecks}

One practical way to achieve accuracy is to scale up the widths of all the layers in the network. The wider they are, the more non-linearities there are in the model, and in practice this improves model accuracy. On the other hand, the size of the matrices involved in the loss and gradient calculations increases, making the underlying \textbf{matmul} computations slower. Unfortunately, the cost of matmul operations (naively) scale up quadratically in the size of their inputs. To compute the output of a hidden layer $H_i = \sigma(H_{i-1} W_i)$ where $W_i \in \mathbb{R}^{m \times n}$, we perform  $m\times n$ multiply-add operations for each input row in $H_{i-1}$. The `wider is better' strategy typically isn't cost-effective \cite{denil13predicting}. We find that carefully inserting \textbf{bottleneck layers} of low-rank matrices between layers of non-linearities greatly reduces scaling costs, with only a small loss of relative accuracy.

Applying singular value decomposition to $W_i$'s, we often observe that the top half of singular values contribute to over $90\%$ of the norm of singular values. This suggests that we can approximate $H_{i-1} W_i$ by a bottleneck layer $H_{i-1} U_iV_i$, where $U_i \in \mathbb{R}^{m \times k}, V_i \in \mathbb{R}^{k \times n}$. The amount of compute reduces to $m\times k + k \times n$, which can be significant for small enough $k$. For a fixed $k$, if we scale $m,n$ by constant $c$, compute scales only linearly with $c$. Empirically, we found that accuracy loss from this approximation was indeed small. By carefully balancing the following two factors, we were able to leverage bottlenecks to achieve better accuracy without increasing computation cost: (1) increasing layer sizes toward better accuracy, at the cost of more compute, and (2) inserting bottleneck layers to reduce compute, at a small loss of accuracy. Balancing of these two can be done manually or via AutoML techniques (discussed in the next section). A recent manual application of this technique to the model (without AutoML tuning) reduced time per training step by $7\%$ without impacting accuracy (See Table~\ref{tab:efficiency-summary} for a summary of efficiency techniques).

\subsection{AutoML for Efficiency} \label{sec:automl}

To develop an ads CTR prediction model architecture with optimal accuracy/cost trade-off, we typically have to tune the embedding widths of dozens of features and layer widths for each layer in the DNN. Assuming even just a small constant number of options for each such width, the combinatorial search space quickly reaches intractable scales. For industrial-scale models, it is not cost-effective to conduct traditional architecture search with multiple iterations \cite{zoph2018learning, real2019regularized}. We have successfully adopted neural architecture search based on weight sharing \cite{bender2020can} to efficiently explore network configurations (e.g., varying layer width, embedding dimension) to find versions of our model that provide neutral accuracy with decreased training and serving cost.
As illustrated in Figure~\ref{fig:automl}, this is achieved by three components: a weight-sharing network, an RL controller, and constraints.

The weight-sharing network builds a super-network containing all candidate architectures in the search space as sub-networks. In this way, we can train all candidate architectures simultaneously in a single iteration and select a specific architecture by activating part of the super-network with masking. This setup significantly reduces the number of exploration iterations from O(1000) to O(1).

The reinforcement learning controller maintains a sampling distribution, $\theta_{dist}$, over candidate networks. It samples a set of decisions ($d_1,d_2,...$) to activate a sub-network at each training step. We then do a forward pass for the activated sub-network to compute loss and cost. Based on that, we estimate the reward value $R(d_1,d_2,...)$ and conduct a policy gradient update using the REINFORCE algorithm \cite{williams1992simple} as follows:
$$
\theta_{dist} = \theta_{dist} + \alpha_0 \cdot (R(d_1,d_2,...) - \overline R) \cdot \nabla \log P(d_1,d_2,...|\theta_{dist}),
$$
where $\overline R$ denotes the moving average value of the reward and $\alpha_0$ is the learning rate for the reinforcement learning algorithm. Through the update at each training step, the sampling rate of better architectures will gradually increase and the sampling distribution will eventually converge to a promising architecture. We select the architecture with maximum likelihood at the end of the training. Constraints specify how to compute the cost of the activated sub-network, which can typically be done by estimating the number of floating-point operations or running a pre-built hardware-aware neural cost model. The reinforcement learning controller incorporates the provided cost estimate into the reward (e.g., $R = R_{\text{accuracy}} + \gamma \cdot |\text{cost}/\text{target} - 1|$, where $\gamma < 0$) \cite{bender2020can} in order to force the sampling distribution to converge to a cost-constrained point. In order to search for architectures with lower training cost but neutral accuracy, in our system we set up multiple AutoML tasks with different constraint targets (e.g. 85\%/90\%/95\% of the baseline cost) and selected the one with neutral accuracy and smallest training cost. A recent application of this architecture search to the model reduced time per training step by $16\%$ without reducing accuracy.

\subsection{Data Sampling} \label{sec:sampling}

Historical examples of clicks on search ads make up a large dataset that increases substantially every day. The diminishing returns of ever larger datasets dictate that it is not beneficial to retain all the data. The marginal value for improving model quality goes toward zero, and eventually does not justify any extra machine costs for training compute and data storage. Alongside using ML optimization techniques to improve ML efficiency, we also use data sampling to control training costs. Given that training is a single-pass over data in time-order, there are two ways to reduce the training dataset: 1) restricting the time range of data consumed; and 2) sampling the data within that range. Limiting training data to more recent periods is intuitive. As we extend our date range further back in time, the data becomes less relevant to future problems. Within any range, clicked examples are more infrequent and more important to our learning task; so we sample the non-clicked examples to achieve rough class balance. 
Since this is primarily for efficiency, exact class balance is unnecessary. A constant sampling rate (a constant class imbalance prior) can be used with a simple single-pass filter. To keep model predictions unbiased, importance weighting is used to up-weight negative examples by the inverse of the sampling rate.
Two additional sampling strategies that have proved effective are as follows:
\begin{itemize}
    \item Sampling examples associated with a low logistic loss (typically examples with low estimated CTR and no click).
    \item Sampling examples that are very unlikely to have been seen by the user based on their position on the page.
\end{itemize}

The thresholds for the conditions above are hand-tuned and chosen to maximize data reduction without hurting model accuracy. These strategies are implemented by applying a small, constant sampling rate to all examples meeting any of the conditions above. Pseudo-Random sampling determines whether examples should be kept and re-weighted or simply discarded. This ensures that all training models train on the same data. This scheme may be viewed as a practical version of \cite{fithian2014local} for large problem instances with expensive evaluation. Simple random sampling allows us to keep model estimates unbiased with simple constant importance re-weighting. It is important to avoid very small sampling rates in this scheme, the consequent large up-weighting can lead to model instability. Re-weighting is particularly important for maintaining calibration, since these sampling strategies are directly correlated to labels. 

For sampling strategies that involve knowing the loss on an example, calculating that loss would require running inference on the training example, removing most of the performance gains. For this reason, we use a proxy value based on a prediction made by a "teacher model". In this two-pass approach. We first train once over all data to compute losses and associated sampling rates, and then once on the sub-sampled data. The first pass uses the same teacher model for distillation (Section \ref{sec:distillation}) and is only done once. Iterative research can then be performed solely on the sub-sampled data. While these latter models will have different losses per example, the first pass loss-estimates still provide a good signal for the `difficulty' of the training example and leads to good results in practice. Overall our combination of class re-balancing and loss-based sampling strategies reduces the data to < 25\% of the original dataset for any given period without significant loss in accuracy.

\section{Accuracy} \label{sec:accuracy}

Next we detail a set of techniques aimed at improving the accuracy of the system. We discuss: additional losses that better align offline training-time metrics with important business metrics, the application of distillation to our online training setting, the adaptation of the Shampoo second-order optimizer to our model, and the use of Deep and \& Cross networks. 

\subsection{Loss Engineering}

Loss engineering plays an important role in our system. As the goal of our model is to predict whether an ad will be clicked, our model generally optimizes for logistic loss, often thought of as the cross-entropy between model predictions and the binary task (click/no-click) labels for each example. Using logistic loss allows model predictions to be unbiased so that the prediction can be interpreted directly as a calibrated probability. Binary predictions can be improved by introducing soft prediction through distillation methods \cite{hinton2017}. 
Beyond estimating the CTR per ad, it is important that the set of candidate ads for a particular query is correctly ranked (such that ads with clicks have higher CTR than ads without clicks), thus incorporating proper ranking losses is also important. In this section, we discuss novel auxiliary losses and introduce multi-task and multi-objective methods for joint training with these losses 

\subsubsection{Rank Losses} \label{sec:rank_loss}
We found that Area under the ROC curve computed per query (PerQueryAUC) is a metric well correlated with business metrics quantifying the overall performance of a model. In addition to using PerQueryAUC during evaluation, we also use a relaxation of this metric, i.e., rank-loss, as a second training loss in our model. There are many rank losses in the learning-to-rank family \cite{pasumarthi2019, burges2010ranknet}. 
We find one effective approximation is Ranknet loss \cite{burges2005ltr}, which is a pairwise logistic loss:
$$
-\sum_{i \in \{y_i = 1\}}\sum_{j \in \{y_j \neq 1\}} \log(\text{sigmoid}(s_i, s_j)),
$$
where $s_i, s_j$ are logit scores of two examples.

Rank losses should be trained jointly with logistic loss; there are several potential optimization setups. In one setup, we create a multi-objective optimization problem \cite{sculley10combinedregression}:
$$
\mathcal{L}(W) = \alpha_1 \mathcal{L}_\text{rank}(\mathbf{y}_\text{rank}, \mathbf{s}) + (1 - \alpha_1)\mathcal{L}_\text{logistic}(\mathbf{y}, \mathbf{s}),  
$$
where $\mathbf{s}$ are logit scores for examples, $\mathbf{y}_\text{rank}$ are ranking labels, $\mathbf{y}$ are the binary task labels, and $\alpha_1 \in (0, 1)$ is the rank-loss weight. Another solution is to use multi-task learning \cite{caruana1997mtl, ruder2017overview}, where the model produces multiple different estimates $s$ for each loss.
\begin{align*}
& \mathcal{L}(W_{\text{shared}}, W_{\text{logistic}}, W_{\text{rank}}) = \\
& \qquad \alpha_1 \mathcal{L}_\text{rank}(\mathbf{y}, \mathbf{s}_\text{rank}) + (1 - \alpha_1)\mathcal{L}_\text{logistic}(\mathbf{y}, \mathbf{s}_\text{logistic}),
\end{align*}
where $W_{\text{shared}}$ are weights shared between the two losses, $W_{\text{logistic}}$ are for the logistic loss output, and $W_{\text{rank}}$ are for the rank-loss output. In this case, the ranking loss affects the "main" prediction $\mathbf{s}_\text{logistic}$ as a "regularizer" on $W_\text{shared}$.

As rank losses are not naturally calibrated predictors of click probabilities, the model's predictions will be biased. A strong bias correction component is needed to ensure the model's prediction is unbiased per example. More detail can be found in Section~\ref{sec:copt}. Application of ranklosses to the model generated accuracy improvements of $-0.81\%$ with a slight increase in training cost of $1\%$.

\subsubsection{Distillation.}  \label{sec:distillation}
Distillation adds an additional auxiliary loss requiring matching the predictions of a high-capacity teacher model, treating teacher predictions as soft labels \cite{hinton2017}. 
In our model, we use a \textbf{two-pass online distillation} setup. On the first pass, a teacher model records its predictions progressively before training on examples. Student models consume the teacher's predictions while training on the second pass. Thus, the cost of generating the predictions from the single teacher can be amortized across many students (without requiring the teacher to repeat inference to generate predictions). In addition to improving accuracy, distillation can also be used for reducing training data costs. Since the high-capacity teacher is trained once, it can be trained on a larger data set. Students benefit implicitly from the teachers prior knowledge of the larger training set, and so require training only smaller and more recent data. The addition of distillation to the model improved accuracy by $~0.41\%$ without increasing training costs (in the student).

\subsubsection{Curriculums of Losses}
In machine learning, curriculum learning \cite{Bengio2009Curriculum} typically involves a model learning easy tasks first and gradually switching to harder tasks. We found that training on all classes of losses in the beginning of training increased model instability (manifesting as outlier gradients which cause quality to diverge). Thus, we apply an approach similar to curriculum learning to ramp up losses, starting with the binary logistic loss and gradually ramping up distillation and rank losses over the course of training.

\begin{figure*}[t]
\begin{center}
\vspace{-0.25cm}
\includegraphics[width=0.9\textwidth]{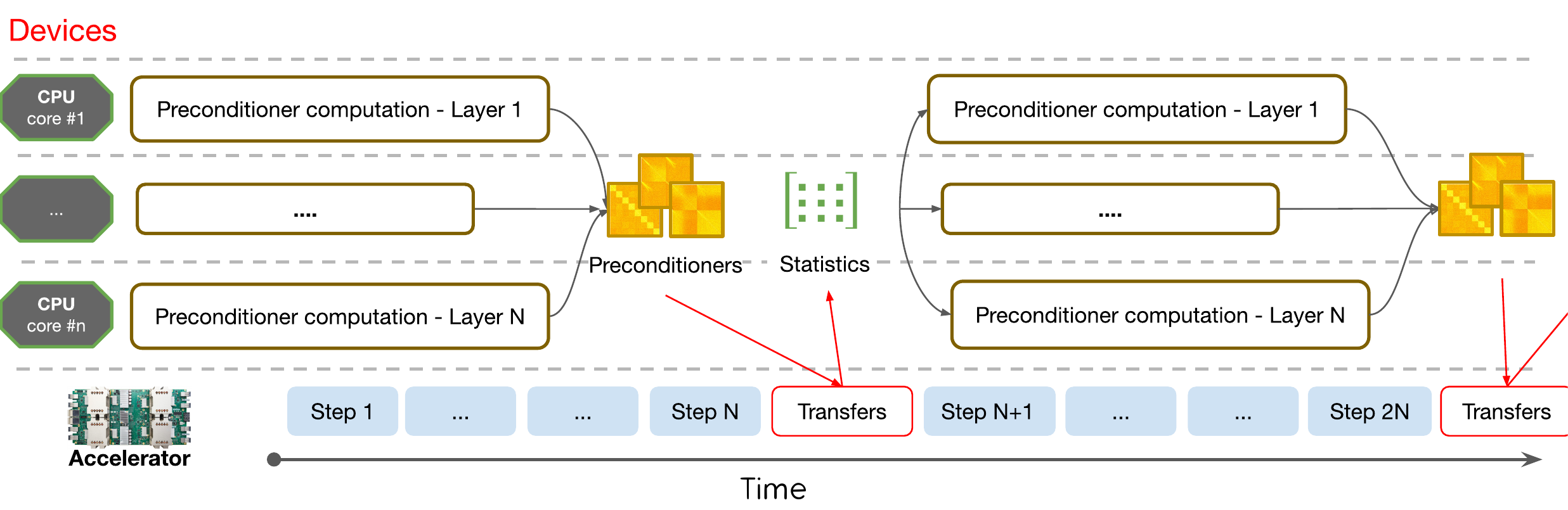}
\vspace{-0.5cm}
\end{center}
\caption{\small Distributed Shampoo \citep{anil2020shampoo}: inverse-$p^{th}$ root computations in double precision runs every $N$ steps and asynchronously pipelined on all CPU cores attached to the TPU accelerators.	}
\label{fig:system}
\end{figure*}

\subsection{Second-order Optimization} \label{sec:secondorder}
Second-order optimization methods that use second derivatives or second-order statistics are known to have better convergence properties compared to first-order methods \cite{NoceWrig06}. Yet to our knowledge, second-order methods are rarely reported to be used in production ML systems for DNNs. Recent work on \mbox{Distributed Shampoo} \cite{anil2020shampoo, gupta2018shampoo} has made second-order optimization feasible for our model by leveraging the heterogeneous compute offered by TPUs and host-CPUs, and by employing additional algorithmic efficiency improvements.

For our model, \mbox{Distributed Shampoo} provided much faster convergence with respect to training steps, and yielded better accuracy when compared to standard adaptive optimization techniques including AdaGrad \cite{duchi2011adaptive}, Adam \cite{kingma2014adam}, Yogi \cite{zaheer2018adaptive}, and LAMB \cite{you2019large}. While second-order methods are known to provide faster convergence compared to first-order methods in the literature - It often fails to provide competitive wall-clock time due to the computational overheads in the optimizer, especially on smaller scale benchmarks. For our model, second-order optimization method was an ideal candidate due to the large batch sizes used in training which amortizes the cost of costly update rule. Training time only increased by approximately 10\%, and the improvements to model accuracy far outweighed the increase in training time. We next discuss salient implementation details specific to our model. 

\emph{Learning Rate Grafting.} 
One of the main challenges in online optimization is defining a learning rate schedule. In contrast to training on static datasets, the number of steps an online model will require is unknown and may be unbounded. Accordingly, popular learning rate schedules from literature depending on fixed time horizons, such as cosine decay or exponential decay, perform worse in contrast to the implicit data-dependent adaptive schedule from AdaGrad \cite{duchi2011adaptive}. 
As observed in literature \cite{agarwal2020disentangling}, 
we also find that AdaGrad's implicit schedule works quite well in the online setting; especially after the $\epsilon$ parameter (the initial accumulator value) is tuned. Accordingly, we bootstrap the schedule for Distributed Shampoo via grafting the per-layer step size from AdaGrad.
More precisely, we use the direction from Shampoo while using the magnitude of step size from AdaGrad at a per-layer granularity. An essential feature of this bootstrapping is that it allowed us to inherit hyper-parameters from previous AdaGrad tunings to search for a Pareto optimal configuration. 

\emph{Momentum.} 
Another effective implementation choice is the combination of Nesterov-styled momentum with the preconditioned gradient. Our analysis suggests that momentum added modest gains on top of Shampoo without increasing the computational overhead while marginally increasing the memory overhead. Computational overhead was addressed via the approximations described in \cite{pmlr-v28-sutskever13}. 

\emph{Stability \& Efficiency.} 
Distributed Shampoo has higher computational complexity per step as it involves matrix multiplication of large matrices for preconditioning and statistics/preconditioner computation. We addressed these overheads with several techniques in our deployment. For example, the block-diagonalization suggested in \cite{anil2020shampoo} effectively reduced computational complexity while also allowing the implementation of parallel updates for each block in the data-parallel setting via weight-update sharding \cite{xu2020automatic}. This optimization reduced the overall step time. Moreover, optimizer overheads are independent of batch size; thus, our use of large batch sizes helped reduce overall computational overhead. Finally, we found that the condition number of statistics used for preconditioning can vary in range, reaching more than $10^{10}$. As numerical stability and robustness are of utmost importance in production, we use double precision numerics. To compute the preconditioners, we use the CPUs attached to the TPUs to run inverse-$p$th roots and exploit a faster algorithm; the coupled Newton iteration \cite{guo2006schur} for larger preconditioners as in Figure~\ref{fig:system}. 

When integrated with the ad click prediction model, the optimizer improved our primary measure of accuracy, Area under the ROC curve computed per query (PerQueryAuc), by $0.44\%$. Accuracy improvements above 0.1\% are considered significant. For comparison: a naive scaling of the deep network by 2x yields a PerQueryAUC improvement of $0.13\%$. See Table~\ref{tab:accuracy-summary} for a summary of accuracy technique results.

\begin{table*}[h]
\parbox{.6\linewidth}{
\begin{tabular}{lSSS} \toprule
    \multicolumn{1}{m{3cm}}{\centering Technique} & 
    \multicolumn{1}{p{1.7cm}}{\centering Accuracy \\ Improvement} &
    \multicolumn{1}{p{1.8cm}}{\centering Training Cost \\ Increase} & 
    \multicolumn{1}{p{1.9cm}}{\centering Inference Cost \\ Increase} \\ \midrule
    {Deep \& Cross Network} & {0.18\%} & {3\%} & {1\%} \\
     Distributed Shampoo Optimizer & {0.44\%} & {10\%} & {0\%} \\
     Distillation & {0.46\%}  & {<1\%$^*$} & {0\%} \\
     Rank Losses & {0.81\%}  & {<1\%} & {0\%} \\ \midrule
     {Baseline: 2x DNN Size} & {0.13\%}  & {36\%} & {10\%} \\
     \bottomrule
\end{tabular}
\vspace{0.2cm}
\caption{\label{tab:accuracy-summary} Accuracy improvement and training/inference costs for accuracy improving techniques. {$^*$} Distillation does not include teacher cost which, due to amortization, is a small fraction of overall training costs.}
}
\hfill
\parbox{.3\linewidth}{
\centering
\begin{tabular}{lS} \toprule
    \multicolumn{1}{m{1.8cm}}{\centering Technique} & 
    \multicolumn{1}{m{1.8cm}}{\centering Training Cost Decrease}  \\ \midrule
    {Bottlenecks} & {7\%} \\
    {AutoML} &{16\%} \\
    {Data Sampling} & {75\%} \\
     \bottomrule
\end{tabular}
\vspace{0.2cm}
\caption{\label{tab:efficiency-summary} Training cost improvements of applied techniques. }
}
\end{table*}
\subsection{Deep \& Cross Network} \label{sec:dcn}
Learning effective feature crosses is critical for recommender systems \cite{zhou2018deep, dcnv2}. We adopt an efficient variant of DCNv2 \cite{dcnv2} using bottlenecks. This is added between the embedding layer $e$ described in Section~\ref{sec:systems} and the DNN. We next describe the Deep \& Cross Network architecture and its embedding layer input. We use a standard embedding projection layer for sparse categorical features. We project categorical feature $i$ from a higher dimensional sparse space to a lower dimensional dense space using $\tilde{e}_i = W_i x_i$, where $x_i \in \{0, 1\}^{v_i}$; $W_i \in \mathbb{R}^{m_i \times v_i}$ is the learned projection matrix; $\tilde{e}_i$ is the dense embedding representation; and $v_i$ and $m_i$ represent the vocabulary and dense embedding sizes respectively. For multivalent features, we use average pooling of embedding vectors. Embedding dimensions $\{m_i\}$ are tuned for efficiency and accuracy trade-offs using AutoML (Section \ref{sec:automl}). Output of the embedding layer is a wide concatenated vector $e_0 = \text{concat}(\tilde{e}_1, \tilde{e}_2 \ldots \tilde{e}_F) \in \mathbb{R}^m$ for $F$ features. For crosses, we adopt an efficient variant of \cite{dcnv2}, applied directly on top of the embedding layer to explicitly learn feature crosses:
$ e_{i} = \alpha_2 \big( e_0 \odot U_i V_i e_{i-1} \big) + e_{i-1}$,
where $e_i, e_{i-1} \in \mathbb{R}^m$ represent the output and input of the $i^\text{th}$ cross layer, respectively; $U_i\in \mathbb{R}^{m \times k}$ and $V_i \in \mathbb{R}^{k \times m}$ are the learned weight matrices leveraging bottlenecks (Section \ref{sec:bottlenecks}) for efficiency; $\alpha_2$ is a scalar, ramping up from $0 \rightarrow 1$ during initial training, allowing the model to first learn the embeddings and then the crosses in a curriculum fashion. Furthermore, this ReZero initialization \cite{rezero} also improves model stability and reproducibility (Section~\ref{sec:reproducibility}).

In practice adding the Deep \& Cross Network to the model yielded an accuracy improvement of $~0.18\%$ with a minimal increase in training cost of $3\%$.

\subsection{Summary of Efficiency and Accuracy Results}
\label{sec:summary-eff-acc}

Below we share measurements of the relative impact of the previously discussed efficiency and accuracy techniques as applied to the production model. The goal is to give a very rough sense of the impact of these techniques and their accuracy vs. efficiency tradeoffs. While precise measures of accuracy improvement on one particular model are not necessarily meaningful, we believe the coarse ranking of techniques and rough magnitude of results are interesting (and are consistent with our general experience).

The baseline 2x DNN size model doubles the number of hidden layers. Note, that sparse embedding lookups add to the overall training cost, thus doubling the number layers does not proportionally increase the cost.

\section{Irreproducibility}
\label{sec:reproducibility}
Irreproducibility, noted in Section~\ref{sec:intro}, may not be easy to detect because it may appear in post deployment {\bf system metrics} and not in progressive validation quality metrics. 
A pair of duplicate models may converge to two different optima of the highly non-convex objective, giving equal average accuracy, but different individual predictions, but with different downstream system/auction outcomes. Model deployment leads to further divergence, as ads selected by deployed models become part of subsequent training examples \cite{swaminathan15batch}. This can critically affect R\&D: experimental models may appear beneficial, but gains may disappear when they are retrained and deployed in production.
Theoretical analysis is complex even in the simple convex case, which is considered only in very recent work \cite{ahn21reproducibility}.
Many factors contribute to irreproducibility \cite{fort2020deep,frankle2020linear,shallue18,snapp2021synthesizing,summers21,zhuang2021randomness}, including random initialization, non-determinism in training due to highly-parallelized and highly-distributed training pipelines, numerical errors, hardware, and more. Slight deviations early in training may lead to very different models \cite{achille17}.
While standard training metrics do not expose system irreproducibility, we can use deviations of predictions on individual examples as a cheap proxy, allowing us to fail fast prior to evaluation at deployment-time. Common statistical metrics (standard deviation, various divergences) can be used \cite{chen20,yu2021dropout} but they require training many more models, which is undesirable at our scale. Instead, we use the \textbf{Relative Prediction Difference} (PD) \cite{shamir20a,shamir2020smooth} metric
$$\Delta_r \stackrel{\triangle}{=} 1/M \cdot \sum_i |\hat{y}_{i,1} - \hat{y}_{i,2}| / [(\hat{y}_{i,1} + \hat{y}_{i,2}) / 2]$$, 
measuring absolute point-wise difference in model predictions for a pair of models (subscripts $1$ and $2$), normalized by the pair's average prediction. Computing PD requires training a pair of models instead of one, but we have observed that reducing PD is sufficient to improve reproducibility of important system metrics. In this section, we focus on methods to improve PD; Section~\ref{sec:copt} focuses on directly improving system metrics.

PDs may be as high as $20\%$ for deep models. 
Perhaps surprisingly, standard methods such as fixed initialization, regularization, dropout, data augmentation, as well as new methods imposing constraints \citep{bho21,shamir18} either failed to improve PD or improved PD at the cost of accuracy degradation. Techniques like warm-starting model weights to values of previously trained models may not be preferable because they can anchor the model to a potentially bad solution space and do not help the development cycle for newer more reproducible models for which there is no anchor.

Other techniques have shown varying levels of success. Ensembles~\cite{dietterich00}, specifically {\em self}-ensembles \cite{allen2020towards}, where we average predictions of multiple model duplicates (each initialized differently), can reduce prediction variance and PD. However, maintaining ensembles in a production system with multiple components builds up substantial technical debt \cite{sculley2014machine}. While some literature \cite{kondratyuk2020ensembling,lobacheva2020power,wang2021wisdom} describes accuracy advantages for ensembles, in our regime, ensembles degraded accuracy relative to equal-cost single networks. 
We believe this is because, unlike in the benchmark image models,
examples in online CTR systems are visited once, and, more importantly, the learned model parameters are dominated by sparse embeddings. Relatedly, more sophisticated techniques based on ensembling and constraints can also improve PD \cite{anil2018large,shamir20a,shamir20}.

Techniques described above trade accuracy and complexity for better reproducibility, requiring either ensembles or constraints. Further study and experimentation revealed that the popular use of Rectified Linear Unit (ReLU) activations contributes to increased PD.
ReLU's gradient discontinuity at 0 induces a highly non-convex loss landscape. Smoother activations, on the other hand, reduce the amount of non-convexity, and can lead to more reproducible models \cite{shamir2020smooth}.
Empirical evaluations of various smooth activations \cite{barron17,hendrycks16,ramachandran17,zheng15} have shown not only better reproducibility compared to ReLU, but also slightly better accuracy. The best reproducibility-accuracy trade-offs in our system were attained by the simple \emph{Smooth reLU (SmeLU)} activation proposed in \cite{shamir2020smooth}. The function form is: 

\begin{equation}
 \label{eq:smelu}
 f_{{}_\text{SmeLU}}(z) = \left \{
   \begin{array}{rl}
  0; & z \leq -\beta \\
  \frac{(z + \beta)^2}{4\beta}; & |z| \leq \beta \\
     z; & z \geq \beta.
  \end{array}
  \right .
 \end{equation}

In our system, $3$-component ensembles reduced PD from 17\% to 12\% and anti-distillation reduced PD further to 10\% with no accuracy loss. SmeLU allowed launching a non-ensemble model with PD less than 10\% that also improved accuracy by 0.1\%.
System reproducibility metrics also improved to acceptable levels compared to the unacceptable levels of ReLU single component models.

\section{Generalizing Across UI Treatments}
\label{sec:uinorms}

\begin{figure}
\centering
  \includegraphics[width=0.5\textwidth]{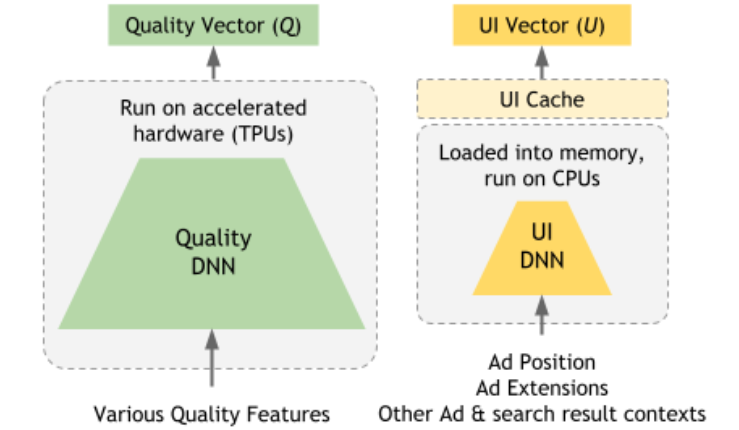}
  \captionof{figure}{\small Model factorization into separable Quality and UI models with estimated CTR $:= \tau(Q \cdot U)$}
  \label{fig:fvm}
\end{figure}

One of the major factors in CTR performance of an ad is its \textbf{UI treatment}, including positioning, placement relative to other results on the page, and specific renderings such as bolded text or inlined images. A complex auction must explore not just the set of results to show, but how they should be positioned relative to other content, and how they should be individually rendered \cite{cavallo17sponsored}. This exploration must take place efficiently over a combinatorially large space of possible treatments.

We solve this through model factorization, replacing estimated CTR with $\tau(Q \cdot U)$, composed of a transfer function $\tau$ where $Q$, $U$ are separable models that output \emph{{vectorized}} representations of the \emph{{Quality}} and the \emph{{UI}}, respectively, and are combined using an inner-product. While $Q$, consisting of a large DNN and various feature embeddings, is a costly model, it needs to be evaluated \emph{{only once}} per ad, irrespective of the number of UI treatments. In contrast, $U$, being a much lighter model, can be evaluated hundreds of times per ad. Moreover, due to the relatively small feature space of the UI model, outputs can be cached to absorb a significant portion of lookup costs (as seen in Figure \ref{fig:fvm}).

Separately from model performance requirements, accounting for the influence of UI treatments on CTR is also a crucial factor for model quality. Auction dynamics deliberately create strong correlations between individual ads and specific UI treatments. Results that are lower on the page may have low CTR regardless of their relevance to the query. Failure to properly disentangle these correlations creates inaccuracy when generalizing over UI treatments (e.g., estimating CTR if the same ad was shown higher on the page). Pricing and eligibility decisions depend crucially on CTR estimates of sub-optimal UIs that are rarely occurring in the wild. For instance, our system shouldn't show irrelevant ads, and so such scenarios will not be in the training corpus, and so estimates of their irrelevance (low CTR) will be out of distribution. But these estimates are needed to ensure the ads do not show. Even for relevant ads, there is a similar problem. Performance of ads that rarely show in first position may still be used to set the price of those ads that often do show in first position. This creates a specific generalization problem related to UI, addressed in Section \ref{sec:copt}. 

Calibration is an important characteristic for large-scale ads recommendation. We define calibration bias as label minus prediction, and want this to be near zero \emph{{per ad}}. 
A calibrated model allows us to use estimated CTR to determine the trade-off between showing and not showing an ad, and between showing one ad versus another; both calculations can be used in downstream tasks such as UI treatment selection, auction pricing, or understanding of ad viewability.

\begin{figure*}[h]
  \centering
  \includegraphics[width=0.85\textwidth]{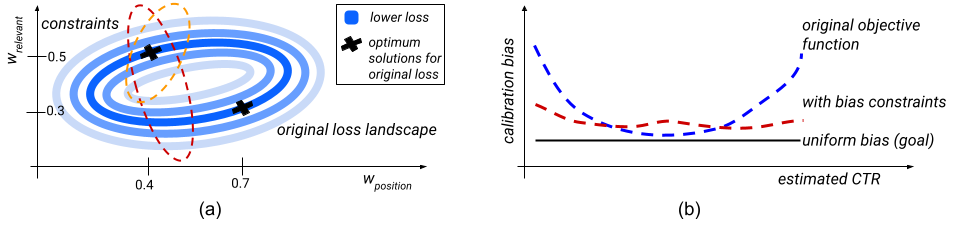}
  \captionof{figure}{\small (a) Loss landscape for a model with non-identifiability across two weights and how bias constraints help find the right solution: we add additional criteria (red and orange curves) such that we choose the correct solution at optimum loss (dark blue curve). (b) Calibration bias across buckets of estimated CTR. For calibrated predictions, we expect uniform bias (black curve). Whereas a model with the original objective function is biased for certain buckets of estimated CTR (blue curve), we can get much closer to uniform with bias constraints (red curve).}
  \vspace{-0.25cm}
  \label{fig:copt}
\end{figure*}

The related concept of \textbf{{credit attribution}} is similar to counterfactual reasoning \cite{bottou2013counterfactual} or bias in implicit feedback \cite{joachims2017}. It is a specific non-identifiability in model weights that can contribute to irreproducibility (Section \ref{sec:reproducibility}). Consider an example to illustrate the UI effect (Section~\ref{sec:uinorms}): assume that model A has seen many training examples with high-CTR ads in high positions, and (incorrectly) learned that ad position most influences CTR. Model B, defined similar to A, trains first on the few examples where high-CTR ads appear in low positions, and (correctly) learns that something else (e.g., ad relevancy to query) is causing high CTR. Both models produce the same estimated CTR for these ads but for different reasons, and when they are deployed, model A will likely show fewer ads because it will not consider otherwise useful ads in lower positions; these models will show system irreproducibility.

In our system, we use a novel, general-purpose technique called \textbf{{bias constraints}} to address both calibration and credit attribution. We add calibration bias constraints to our objective function, enforced on relevant slices of either the training set or a separate, labelled dataset. This allows us reduce non-identifiability by anchoring model loss to a desired part of the solution space (e.g., one that satisfies calibration) (Figure \ref{fig:copt}a).
By extension, we reduce irreproducibility by anchoring a retrained model to the same solution.

Our technique is more lightweight than other methods used for large-scale, online training (counterfactual reasoning \cite{bottou2013counterfactual}, variations of inverse propensity scoring \citep{joachims2017, Lefortier2016LargescaleVO}): in practice, there are fewer parameters to tune, and we simply add an additional term to our objective rather than changing the model structure. To address calibration, \cite{Borisov2018CalibrationAS} adjusts model predictions in a separate calibration step using isotonic regression, a non-parametric method. Our technique does calibration jointly with estimation, and is more similar to methods which consider efficient optimization of complex and augmented objectives (e.g., \citep{eban2017ScalableLO, mann2007}). Using additional constraints on the objective allows us to address a wide range of calibration and credit attribution issues.

\section{Bias Constraints}
\label{sec:copt}

\subsection{Online Optimization of Bias Constraints}

We now optimize our original objective function
with the constraint that $~\forall k \forall {i \in S_k}, (y_i - \hat{y}_i) = 0$.
Here, ${S_k}$ are subsets of the training set which we'd like to be calibrated (e.g., under-represented classes of data) or new training data that we may or may not optimize the original model weights over (e.g., out-of-distribution or off-policy data gathered from either randomized interventions or exploration scavenging \citep{wang16learning, joachims2017, langford08exploration}). To aid optimization, we first transform this into an unconstrained optimization problem by introducing a dual variable $\lambda_{k,i}$ for each constraint and maximizing the Lagrangian relative to the dual variables. Next, instead of enforcing zero bias per example, we ask that the squared average bias across $S_k$ is zero. This reduces the number of dual variables to $\{\lambda_k\}$, and is equivalent to adding an L2 regularization on $\lambda_k$ with a constraint of zero average bias.
For a constant $\alpha_3$ controlling regularization, and tuned via typical hyperparameter tuning techniques (e.g. grid search), our new optimization is:
$$\min_{W} \max_{\lambda_k} \sum_{i} \mathcal{L}(y_i, \hat{y}_i) + \sum_{k=1}^K \sum_{i \in S_k} (\lambda_k (y_i - \hat{y}_i) - \frac{\alpha_3}{2} \lambda_k^2)$$
Any degraded accuracy or stability is mitigated by combinations of the following tunings, ordered by impact: ramping up the bias constraint term, reducing the learning rate on $\{\lambda_k\}$, increasing $\alpha_3$, or adding more or finer-grained constraints (breaking up $S_k$). We believe the first two can help normalize any differences between the magnitude of the dual variables and other weights, and the latter two help lessen the strength of the bias term if $S_k$ aren't optimally selected.

\subsection{Bias Constraints for General Calibration}

If we plot calibration bias across buckets of interesting variables, such as estimated CTR or other system metrics, we expect a calibrated model to have uniform bias. However, for several axes of interest, our system shows higher bias at the ends of the range (Figure \ref{fig:copt}b). We apply bias constraints to this problem by defining $S_k$ to be examples in each bucket of, e.g., estimated CTR. Since we don't use the dual variables during inference, we can include estimated CTR in our training objective. With bias constraints, bias across buckets of interest becomes much more uniform: variance is reduced by more than half. This can in turn improve accuracy of downstream consumers of estimated CTR.

\subsection{Exploratory Data and Bias Constraints}

We can also use bias constraints to solve credit attribution for UI treatments. We pick $S_{k}$ by focusing on classes of examples that represent uncommon UI presentations for competitive queries where the ads shown may be quite different. For example, $S_1$ might be examples where a high-CTR ad showed at the bottom of the page, $S_2$ examples where a high-CTR ad showed in the second-to-last position on the page, etc. Depending on how model training is implemented, it may be easier to define $S_{k}$ in terms of existing model features (e.g., for a binary feature ${f}$, we split one sum over $S_k$ into two sums). We choose $\{f\}$ to include features that generate partitions large enough to not impact convergence but small enough that we expect the bias per individual example will be driven to zero (e.g., if we think that query language impacts ad placement, we will include it in $\{f\}$). For the model in Table \ref{table:constraints}, we saw substantial bias improvements on several data subsets $S_k$ related to out-of-distribution ad placement and more reproducibility with minimal accuracy impact when adding bias constraints.

\begin{table}[b]
    \begin{tabular}{|c|c|c|c|c|}
        \hline
        $S_1$ Bias & $S_2$ Bias & $S_3$ Bias & Loss & Ads/Query Churn\\
        \hline
        -15\% & -75\% & -43\% & +0.03\% & -85\% \\
        \hline
    \end{tabular}
\vspace{0.25cm}
    \caption{\small Progressive validation and deployed system metrics reported as a percent change for a bias constraint over the original model (negative is better). Ads/Query Churn records how much the percent difference in the number of ads shown above search results per query between two model retrains changes when deployed in similar conditions; we want this to be close to zero.}
    \label{table:constraints}
\vspace{-0.5cm}
\end{table}

Viewing the bias constraints as anchoring loss rather than changing the loss landscape (Figure \ref{fig:copt}a), we find that the technique does not fix model irreproducibility but rather mitigates system irreproducibility: we were able to cut the number of components in the ensemble by half and achieve the same level of reproducibility.

 \section{Conclusion}
We detailed a set of techniques for large-scale CTR prediction that have proven to be truly effective ``in production'': balancing improvements to accuracy, training and deployment cost, system reproducibility and model complexity---along with describing approaches for generalizing across UI treatments. We hope that this brief visit to the factory floor will be of interest to ML practitioners of CTR prediction systems, recommender systems, online training systems, and more generally to those interested in large industrial settings.

\bibliographystyle{ACM-Reference-Format}
\bibliography{references}

\end{document}